\documentclass{aastex62}

\received{...}
\revised{...}
\accepted{...}

\submitjournal{ApJL}

\shorttitle{Reconnection-condensation filament formation}
\shortauthors{Li et al.}

\begin{document}

\title{Formation of a solar filament by magnetic reconnection and coronal condensation}

\correspondingauthor{Leping Li}
\email{lepingli@nao.cas.cn}

\author[0000-0002-0786-7307]{Leping Li}
\affil{CAS Key Laboratory of Solar Activity, National Astronomical Observatories, Chinese Academy of Sciences, Beijing 100101, People's Republic of China}
\affiliation{University of Chinese Academy of Sciences, Beijing 100049, People's Republic of China}

\author{Hardi Peter}
\author{Lakshmi Pradeep Chitta}
\affiliation{Max Planck Institute for Solar System Research, D-37077 G\"{o}ttingen, Germany}

\author{Hongqiang Song}
\affiliation{Shandong Provincial Key Laboratory of Optical Astronomy and Solar-Terrestrial Environment, and Institute of Space Sciences, Shandong University, Weihai, Shandong 264209, People's Republic of China}

\begin{abstract}

In solar filament formation mechanisms, magnetic reconnection between two sets of sheared arcades forms helical structures of the filament with numerous magnetic dips, and cooling and condensation of plasma trapped inside the helical structures supply mass to the filament.
Although each of these processes, namely, magnetic reconnection and coronal condensation have been separately reported, observations that show the whole process of filament formation are rare.
In this Letter, we present the formation of a sigmoid via reconnection between two sets of coronal loops, and the subsequent formation of a filament through cooling and condensation of plasma inside the newly formed sigmoid.
On 2014 August 27, a set of loops in the active region 12151 reconnected with another set of loops that are located to the east.
A longer twisted sigmoidal structure and a set of shorter lower-lying loops then formed.
The observations coincide well with the tether-cutting model.
The newly formed sigmoid remains stable and does not erupt as a coronal mass ejection.
From the eastern endpoint, signatures of injection of material into the sigmoid (as brightenings) are detected, which closely outline the features of increasing emission measure at these locations. This may indicate the chromospheric evaporation caused by reconnection, supplying heated plasma into the sigmoid.
In the sigmoid, thermal instability occurs, and rapid cooling and condensation of plasma take place, forming a filament.
The condensations then flow bi-directionally to the filament endpoints.
Our results provide a clear observational evidence of the filament formation via magnetic reconnection and coronal condensation.

\end{abstract}


\section{Introduction} \label{sec:int}

Solar filaments (prominences) are composed of cooler and denser chromospheric material suspended in the hotter and rarer corona along magnetic polarity inversion lines (PILs) \citep{2010SSRv..151..333M, 2019A&A...627L...5C}.
They are observed as dark filaments on the disk, and bright prominences above the limb in H$\alpha$ channel \citep{2015ApJS..219...17Y, 2017ApJ...835...94O}.
In the formation of filaments, magnetic reconnection, rearranging the magnetic field topology, is considered to play a key role \citep{2000mrmt.conf.....P}.
Based on reconnection theory, numerous models of filament formation have been developed \citep{pneu83, 1989ApJ...343..971V}. 
In these models, shearing and converging motions of the photospheric magnetic fields, surrounding the PILs, result in reconnection between sheared arcades, forming the helical structures of a filament with numerous dips \citep{1989ApJ...343..971V}.
Subsequently, thermal instability within the helical structures leads to rapid cooling and condensation of plasma, supplying cooler and denser material to the filament \citep{pneu83}.
Here, the magnetic field evolution, e.g., magnetic reconnection, and the thermal evolution, e.g., coronal condensation, are treated separately.
Recently, \citet{kane15, kane17} proposed a reconnection-condensation filament formation model, where reconnection can lead to flux rope formation and also to radiative condensation under some conditions.

Formation of the sigmoid or flux rope and the filament through reconnection has been widely investigated \citep[e.g.,][]{2008ApJ...683..510K, 2014ApJ...797L..15C, 2018ApJ...869...78C, 2019ApJ...874...96Y}.
Employing ultraviolet (UV) and extreme UV (EUV) images, reconnection between two sets of sheared arcades was evidently detected \citep{2012ApJ...753L..29Z, 2014ApJ...797L..15C, 2018ApJ...869...78C}.  
A longer sigmoid or flux rope then formed \citep{2010ApJ...725L..84L, 2015ApJ...812...50J}.
After the appearance, it erupted upward as a coronal mass ejection (CME).
No filament, formed by cooling and condensation of plasma inside the sigmoid or flux rope, was detected  \citep{2010ApJ...725L..84L}.
Using Atmospheric Imaging Assembly \citep[AIA;][]{lemen12} images on board the Solar Dynamics Observatory \citep[SDO;][]{2012SoPh..275....3P}, filament formation through cooling and condensation of coronal plasma has been observed separately in a loop system \citep{liu12} and a cavity \citep{berg12}.
The magnetic structures are, however, pre-existing ones rather than the newly formed ones via reconnection.
Moreover, filament formations through reconnection between two filaments \citep{2014ApJ...795....4J, 2015ApJ...808L..24C, 2017ApJ...838..131Y}, two sets of chromospheric fibrils \citep{2016ApJ...816...41Y, 2017ApJ...840L..23X}, and the filament and loops \citep{2016NatPh..12..847L} were reported.
Subsequently, some of them erupted as CMEs \citep{2014ApJ...795....4J, 2015ApJ...808L..24C}, while others evolved without eruption \citep{2017ApJ...840L..23X, 2016ApJ...816...41Y, 2017ApJ...838..131Y}.
No cooling and condensation of plasma was presented \citep{2016ApJ...816...41Y, 2017ApJ...838..131Y}.

Recently, filament formation through cooling and condensation of plasma in magnetic dips of open structures facilitated by interchange reconnection was found \citep{2018ApJ...864L...4L, 2018ApJ...868L..33L, 2019ApJ...884...34L, 2020ApJ...905...26L, 2021ApJ...910...82L, 2021RAA...under...reviewL}.
In EUV images, the higher-lying open structures move down toward the surface, and reconnect with the lower-lying closed loops \citep{2018ApJ...864L...4L}.
Magnetic dips then form in the former.
Coronal plasma surrounding the dips converges into the dips, resulting in the enhancement of plasma density in the dips.
Triggered by the density enhancement, thermal instability occurs, and rapid cooling and condensation of plasma take place in the dips.
Cool condensation in magnetic dips indicates the formation of a filament \citep{pneu83, karp01, 2010SSRv..151..333M}.
Along with the successive reconnection, the part of open structures, supporting the filament, take part in the reconnection with the lower-lying loops.
Without support from the underlying structures, filament falls to the surface as coronal rain.
It thus represents the transient filament. 
Moreover, repeated formations of the transient filament facilitated by repeated reconnection were reported, indicating that formation of the transient filament, and the subsequent coronal rain, along open structures facilitated by reconnection is common in the corona \citep{2019ApJ...884...34L, 2020ApJ...905...26L, 2021ApJ...910...82L}.

On 2014 August 27, SDO recorded the active region (AR) 12151 in the southern hemisphere, see an Helioseismic and Magnetic Imager \citep[HMI;][]{2012SoPh..275..229S} line-of-sight (LOS) magnetogram in Figure\,\ref{f:general}(c).
In this AR, a sigmoid was formed in the east, see an AIA 171 \AA~image in Figure\,\ref{f:general}(a). 
Subsequently, a filament appeared in the sigmoid, see AIA 304 \AA~and Kanzelh{\"o}he observatory (KSO) H$\alpha$ images separately in Figures\,\ref{f:general}(b) and (d).
In this Letter, we investigate the formation of the sigmoid and the subsequent formation of the filament.
Our results suggest that the sigmoid is formed through reconnection between two sets of loops, and the filament is formed via cooling and condensation of plasma in the newly reconnected sigmoid.
The observations are described in Section\,\ref{sec:obs}. 
The results and a summary and discussion are shown in Sections\,\ref{sec:res} and \ref{sec:sum}, respectively.

\section{Observations}\label{sec:obs}

SDO/AIA is a set of normal-incidence imaging telescopes, acquiring solar atmospheric images in 10 wavelength channels.
Different AIA channels show plasma at different temperatures, e.g., 94\,\AA~peaks at $\sim$7.2 MK (Fe XVIII), 335\,\AA~peaks at $\sim$2.5 MK (Fe XVI), 211 \AA~peaks at $\sim$1.9 MK (Fe XIV), 193 \AA~peaks at $\sim$1.5 MK (Fe XII), 171 \AA~peaks at $\sim$0.9 MK (Fe IX), 131 \AA~peaks at $\sim$0.6 MK (Fe VIII) and $\sim$10 MK (Fe XXI), and 304 \AA~peaks at $\sim$0.05 MK (He II). 
Among them, the 94 \AA~channel is contaminated by warm (mostly around 1 MK) plasma that also contributes to this channel from the 193 and 171 \AA~channels \citep{2012SoPh..275...41B, 2015A&A...583A.109L}. 
In this Letter, we employ AIA 94, 335, 211, 193, 171, 131, and 304 \AA~images, with spatial sampling and time cadence of 0.6\arcsec\,pixel$^{-1}$ and 12 s, to study the formation of the sigmoid and the subsequent filament.
To better show the evolution, the AIA images are enhanced by using the Multiscale Gaussian Normalization (MGN) technique \citep{2014SoPh..289.2945M}.
H$\alpha$ images with spatial sampling of 1\arcsec\,pixel$^{-1}$, provided by the KSO, University of Graz, Austria, are utilized to analyze the evolution of the filament.
HMI LOS magnetograms on board the SDO, with time cadence of 45 s and spatial sampling of 0.5\arcsec\,pixel$^{-1}$, are used to investigate the evolution of associated photospheric magnetic fields.

\section{Results}\label{sec:res}

\begin{figure}[ht!]
\centering
\includegraphics[width=1.\textwidth]{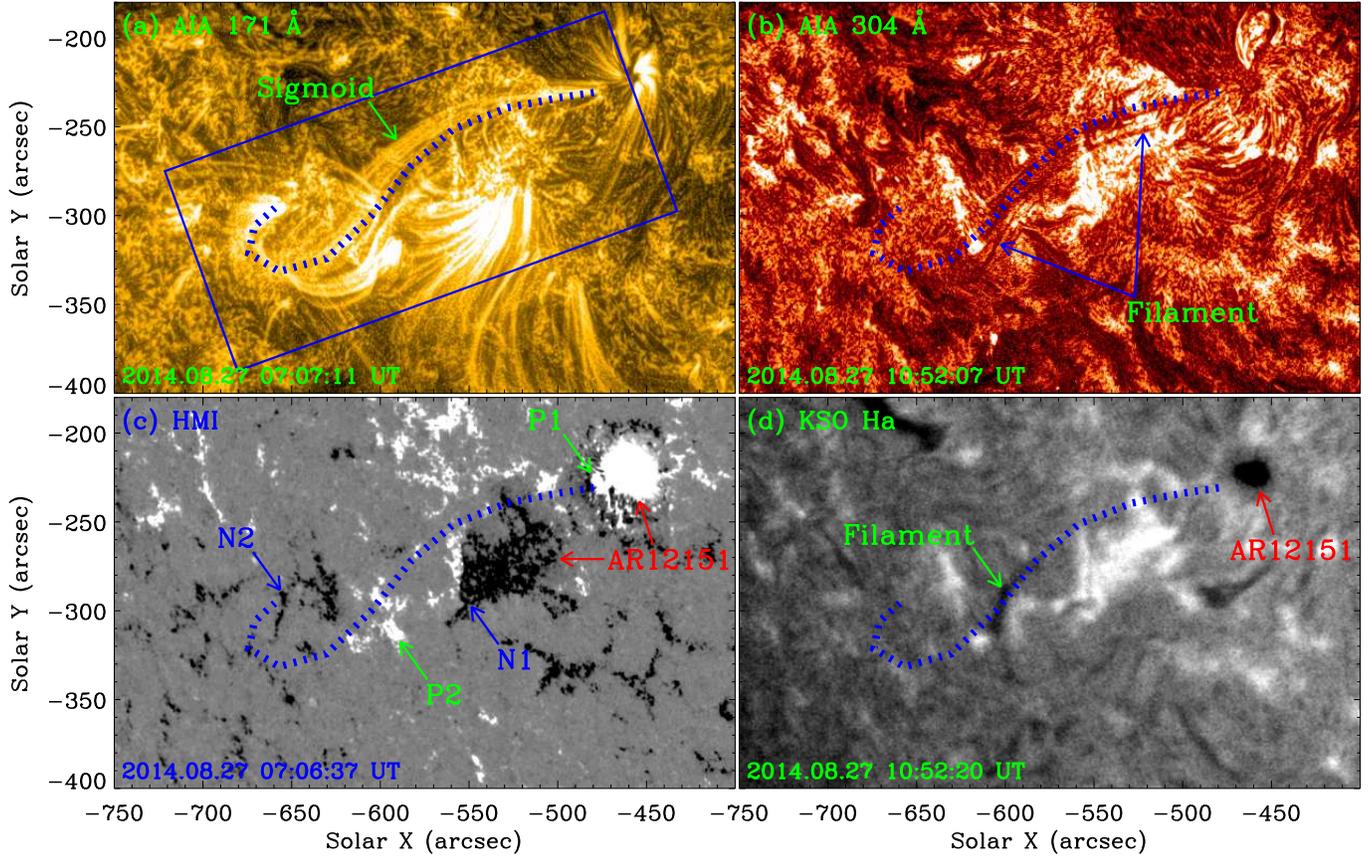}
\caption{General information of the sigmoid and the subsequent filament.
(a) SDO/AIA 171 and (b) 304\,\AA~images, (c) SDO/HMI LOS magnetogram, and (d) KSO H$\alpha$ image. 
Here, the AIA images in (a) and (b) have been enhanced using the MGN technique.
The blue dotted lines outline the sigmoid and the subsequent filament. 
The blue rectangle in (a) denotes the field of view (FOV) of  Figures\,\ref{f:reconnection}, \ref{f:condensations}, and \ref{f:measurements}(a)-(b). 
See Section\,\ref{sec:res} for details. 
\label{f:general}}
\end{figure}

AR 12151 was observed by the SDO at heliographic coordinates S08\,E23 on 2014 August 27, containing a leading sunspot with positive magnetic fields, P1, and the trailing negative magnetic fields, N1, located to the southeast, see Figure\,\ref{f:general}(c).
To the east of the AR, positive and negative magnetic fields, P2 and N2, are detected, see Figure\,\ref{f:general}(c).
From $\sim$06:30 UT, a sigmoid appeared in AIA 171 \AA~images, outlined by the blue dotted line in Figure\,\ref{f:general}(a).
It connects the positive and negative magnetic fields, P1 and N2, see the blue dotted line in Figure\,\ref{f:general}(c).
From $\sim$07:00 UT, a filament in the sigmoid formed in AIA 304 \AA~images, see Figure\,\ref{f:general}(b).
Part of this filament is also observed in H$\alpha$ images, see Figure\,\ref{f:general}(d).

\begin{figure}[ht!]
\includegraphics[width=1.\textwidth]{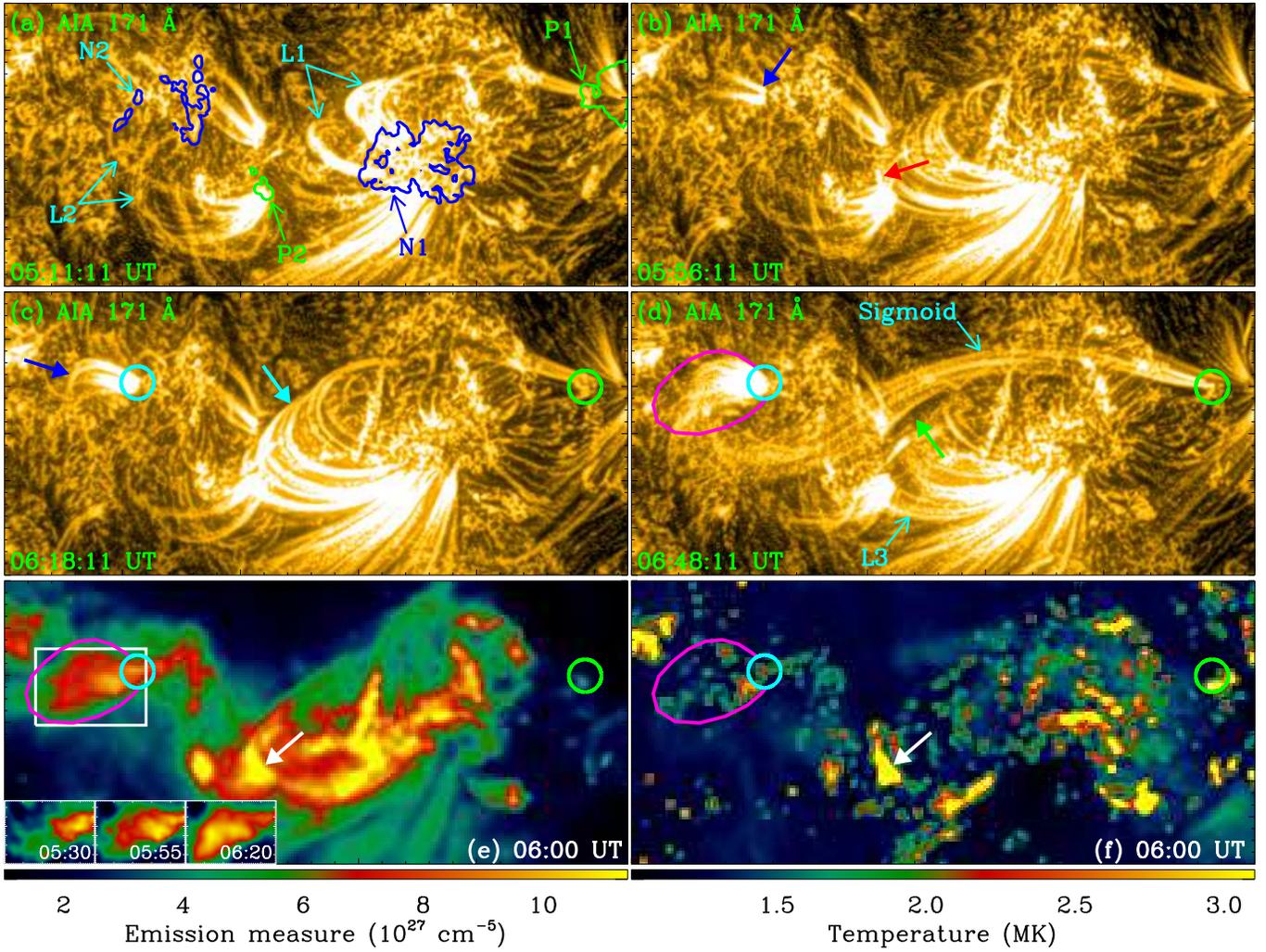}
\centering
\caption{Formation of the sigmoid by reconnection between two sets of loops. 
(a)-(d) AIA 171\,\AA~images enhanced using the MGN technique.
(e) EM and (f) temperature maps obtained using the DEM analysis.
The green and blue contours in (a) enclose positive and negative magnetic fields at the endpoints of loops L1 and L2, respectively.
The blue solid arrows in (b) and (c) mark brightenings at the eastern endpoints of loops L2.
The red solid arrow in (b) shows a bright sheet-like structure.
The cyan and green solid arrows in (c) and (d) separately illustrate the loops L1, and the underlying background after their disappearance. 
The cyan and green circles in (c)-(f) enclose two endpoints of the sigmoid. The pink lines in (d)-(f) enclose region of mass injection into the sigmoid from the eastern endpoint. In (e), the white rectangle shows FOV of the lower-left insets. The white solid arrows in (e) and (f) mark the sheet-like structure.
The FOV is displayed by the blue rectangle in Figure\,\ref{f:general}(a).
See Section\,\ref{sec:res} for details.
\label{f:reconnection}}
\end{figure}

\begin{figure}[ht!]
\includegraphics[width=1.\textwidth]{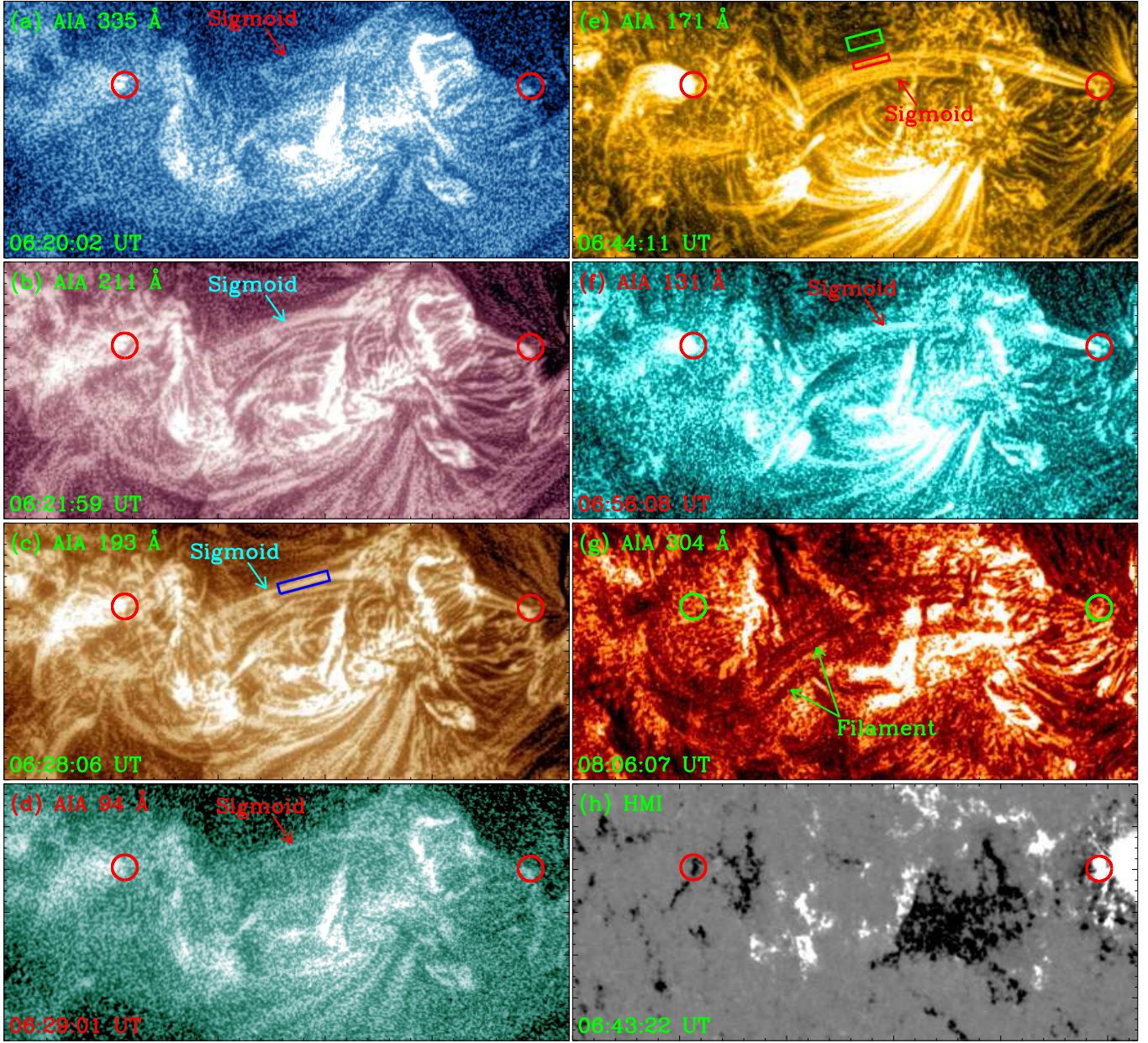}
\centering
\caption{Formation of the filament by cooling and condensation of plasma inside the sigmoid.
(a)-(g) AIA 335, 211, 193, 94, 171, 131, and 304 \AA~images enhanced using the MGN technique, and (h) HMI LOS magnetogram.
The red and green circles enclose two endpoints of the sigmoid and the subsequent filament.
The blue rectangle in (c) marks the region for the light curves of the AIA EUV channels as shown in Figure\,\ref{f:measurements}(d).
The red and green rectangles in (e) separately enclose the region for the DEM curve in Figure\,\ref{f:measurements}(c), and the location where the background emission is measured.
An animation of the unannotated SDO observations is available.
It covers $\sim$9.5 hr starting at 04:30 UT, with time cadence of 1 minute.
The FOV is displayed by the blue rectangle in Figure\,\ref{f:general}(a).
See Section\,\ref{sec:res} for details.
(An animation of this figure is available.)
\label{f:condensations}}
\end{figure}

Before the sigmoid occurred in Figure\,\ref{f:general}(a), two sets of loops, L1 and L2, separately connecting the positive and negative magnetic fields P1 and N1, and P2 and N2, are identified, see Figure\,\ref{f:reconnection}(a).
More loops L1 then appeared, and extended toward the east, see Figure\,\ref{f:reconnection}(b).
A bright sheet-like structure formed at the interface of loops L1 and L2, marked by a red solid arrow in Figure\,\ref{f:reconnection}(b).
Brightenings occurred at the eastern footpoints of loops L2, rooting at the negative magnetic fields N2, and then moved upward along the loops, marked by blue solid arrows in Figures\,\ref{f:reconnection}(b) and (c).
The loops L1, indicated by a cyan solid arrow in Figure\,\ref{f:reconnection}(c), disappeared, revealing the background below, denoted by a green solid arrow in Figure\,\ref{f:reconnection}(d).
Subsequently, a longer higher-lying twisted sigmoid formed, with a length of $\sim$240 Mm, connecting the positive and negative magnetic fields, P1 and N2, see Figure\,\ref{f:general}(c) and green and cyan circles in Figure\,\ref{f:reconnection}(d).
Several threads of the sigmoid are identified, see Figure\,\ref{f:reconnection}(d), with a width of $\sim$2.4 Mm. 
Simultaneously, a set of lower-lying shorter loops, L3, connecting the positive and negative magnetic fields, P2 and N1, appeared, see Figure\,\ref{f:reconnection}(d).
Subsequently, more threads formed along the sigmoid, see the online animated version of Figure\,\ref{f:condensations}.
All the results support that the sigmoid is formed via reconnection between loops L1 and L2.

Using six AIA EUV channels, including 94, 335, 211, 193, 171, and 131\,\AA, we analyze the evolution of emission measure (EM) and temperature of the studied loops. 
Here, we employ the differential EM (DEM) analysis using ``xrt\_dem\_iterative2.pro" \citep{2004IAUS..223..321W, 2012ApJ...761...62C}. 
Figures\,\ref{f:reconnection}(e) and (f) display the EM and temperature maps at 06:00 UT, respectively. 
In these figures, enhancements of the EM and temperature, marked by white solid arrows, are identified at the interface of loops L1 and L2 from $\sim$05:13 UT, where reconnection, lasting for more than 2\,hr, occurred to heat the plasma. 
The EM and temperature also increased at both endpoints of the sigmoid (regions enclosed by cyan and green circles) from $\sim$05:15 UT. 
Moreover, we found that the region near the eastern endpoint exhibited increase in EM (region enclosed by pink lines; see insets in Figure\,\ref{f:reconnection}(e)). 
At the same time, we did not observe a similar development or expansion of region with increasing EM at the site of reconnection. 
This provides evidence for the injection of material into the sigmoid from eastern endpoint after the reconnection.

\begin{figure}[ht!]
\centering
\includegraphics[width=1.\textwidth]{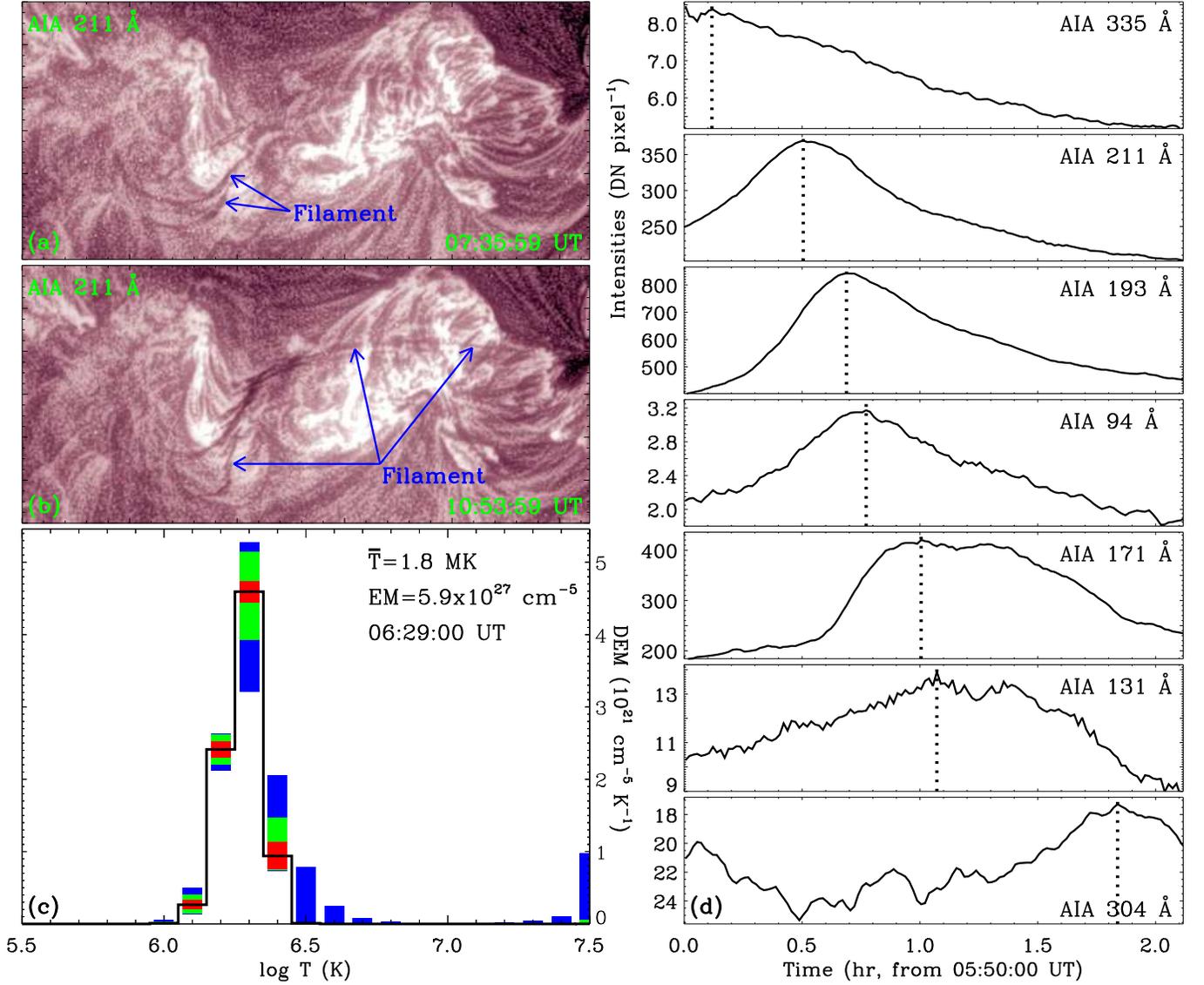}
\caption{Temporal evolution of the sigmoid and the subsequent filament. 
(a) and (b) AIA 211 \AA~images enhanced using the MGN technique.
(c) DEM curve for a sigmoid region enclosed by the red rectangle in Figure\,\ref{f:condensations}(e).  
(d) Light curves of the AIA EUV channels in the blue rectangle in Figure\,\ref{f:condensations}(c). 
Here, the inverse of the AIA 304\,\AA~light curve is displayed.
In (c), the black line is the best-fit DEM distribution, and the red, green, and blue rectangles represent the regions containing 50\%, 51\%-80\%, and 81\%-95\% of the Monte Carlo solutions, respectively.
The dotted lines in (d) separately mark the peaks of the AIA EUV light curves.
The FOV of (a)-(b) is displayed by the blue rectangle in Figure\,\ref{f:general}(a).
See Section\,\ref{sec:res} for details.
\label{f:measurements}}
\end{figure}

The sigmoid first appeared in AIA 335\,\AA~images from $\sim$05:35 UT, and then appeared sequentially in AIA 211, 193, 94, 171, and 131 \AA~images, see Figures\,\ref{f:condensations}(a)-(f).
In AIA 304\,\AA~images, dark condensations formed in the sigmoid since $\sim$06:40 UT, see Figure\,\ref{f:condensations}(g).
Cooling and condensation of plasma in the sigmoid are thus clearly identified.
Condensations in the sigmoid indicate the formation of a filament, rooting at the positive and negative magnetic fields, P1 and N2, denoted by red and green circles in Figure\,\ref{f:condensations}.
Those condensations then flow to the filament endpoints, with speeds of 50-80 km\,s$^{-1}$.
From $\sim$07:10 UT, more condensations formed in the left part of the sigmoid, see Figure\,\ref{f:measurements}(a).
They also move to the filament endpoints bi-directionally, see Figure\,\ref{f:measurements}(b).
The filament remains stable for over a day, and part of it appears in H$\alpha$ images, see Figure\,\ref{f:general}(d).

In the blue rectangle in Figure\,\ref{f:condensations}(c), light curves of the AIA 335, 211, 193, 94, 171, 131, and 304 \AA~channels are calculated, and displayed in Figure\,\ref{f:measurements}(d).
Here, the original, rather than the enhanced, AIA images are employed. As condensations in AIA 304 \AA~images show dark absorption features, the inverse of the AIA 304 \AA~light curve is used. 
Intensity increases in all the AIA EUV light curves, reaches its peak separately at $\sim$05:57 (335\,\AA), $\sim$06:20 (211\,\AA), $\sim$06:31 (193\,\AA), $\sim$06:36 (94\,\AA), $\sim$06:50 (171\,\AA), $\sim$06:54 (131\,\AA), and $\sim$07:40 UT (304\,\AA), see dotted lines in Figure\,\ref{f:measurements}(d), and then decreases.
This also shows the cooling process of the sigmoid evidently.
Before the peak of the AIA 335\,\AA~light curve, no peak is identified in the AIA 94 (131)\,\AA~light curve.
The channel, that would otherwise be sensitive to hotter plasma under flaring conditions, reaches the peak after that of the AIA 193 (171) \AA~light curve.
Thus the sigmoid in the AIA 94 and 131\,\AA~channels, see Figures\,\ref{f:condensations}(d) and (f), is indicative of warm plasma feature.
The sigmoid is heated up to $\sim$2.5 MK, the characteristic temperature of the AIA 335\,\AA~channel, via reconnection.
It then cools down sequentially to $\sim$1.9 MK, the characteristic temperature of the AIA 211\,\AA~channel, in 23 minutes, to $\sim$1.5 MK, the characteristic temperature of the AIA 193\,\AA~channel, in 34 minutes, to $\sim$0.9 MK, the characteristic temperature of the AIA 171\,\AA~channel, in 53 minutes, to $\sim$0.6 MK, the lower characteristic temperature of the AIA 131\,\AA~channel, in 57 minutes, and to $\sim$0.05 MK, the characteristic temperature of the AIA 304\,\AA~channel, in 103 minutes.

When the sigmoid was clearly detected in AIA 94\,\AA~images, see Figure\,\ref{f:condensations}(d), the sigmoid region, enclosed by the red rectangle in Figure\,\ref{f:condensations}(e), is chosen to compute the DEM.
The region out of the sigmoid, enclosed by the green rectangle in Figure\,\ref{f:condensations}(e), is chosen for the background emission that is subtracted from the sigmoid region.
In each region, the DN counts in each of the six AIA EUV channels are averaged spatially over all pixels and normalized temporally by the exposure time.
The DEM curve of the sigmoid region is displayed in Figure\,\ref{f:measurements}(c).
Consistent with the AIA imaging observations, the DEM illustrates a lack of hot plasma component in the sigmoid, see the black curve in Figure\,\ref{f:measurements}(c), supporting that the sigmoid shows warm plasma.
The DEM-weighted temperature and EM are 1.8 MK and 5.9$\times$10$^{27}$ cm$^{-5}$, respectively.
Using the EM, electron number density ($n_{e}$) of the sigmoid is estimated employing $n_{e}=\sqrt{\frac{\mathrm{EM}}{D}}$, where \textit{D} is the sigmoid LOS depth.
Assuming that the depth \textit{D} equals the width (\textit{W}) of the sigmoid, then the density is $n_{e}=\sqrt{\frac{\mathrm{EM}}{W}}$.
Employing EM=5.9$\times$10$^{27}$ cm$^{-5}$ and \textit{W}=2.4 Mm, we obtain the density to be 5$\times$10$^{9}$ cm$^{-3}$.

\section{Summary and discussion}\label{sec:sum}

Employing AIA EUV images and HMI LOS magnetograms, we investigate the formation of a sigmoid and the subsequent filament by reconnection and condensation in AR 12151 on 2014 August 27.
In AIA images, two sets of loops L1 and L2, separately connecting the positive and negative magnetic fields P1 and N1, and P2 and N2, are observed.
At their interface, a bright sheet-like structure appears.
Brightenings and increasing EM then happen at the eastern endpoints of loops L2, and then propagate upward along the loops.
Subsequently, part of the loops L1 disappears, revealing the background below.
A sigmoid and a set of loops L3, respectively connecting the positive and negative magnetic fields P1 and N2, and P2 and N1, then form.
These observations suggest that the sigmoid is formed through reconnection between loops L1 and L2.
The sigmoid appears first in AIA 335\,\AA~images, and then occurs in 211, 193, 94, 171, 131, and 304 \AA~images sequentially. 
The cooling and condensation of plasma, heated up to $\sim$2.5 MK, the characteristic temperature of the AIA 335\,\AA~channel, in the sigmoid is thus clearly detected.
This is also supported by the AIA EUV light curves of the sigmoid.
The condensation in the sigmoid shows the filament formation.

\begin{figure}[ht!]
\centering
\includegraphics[width=0.8\textwidth]{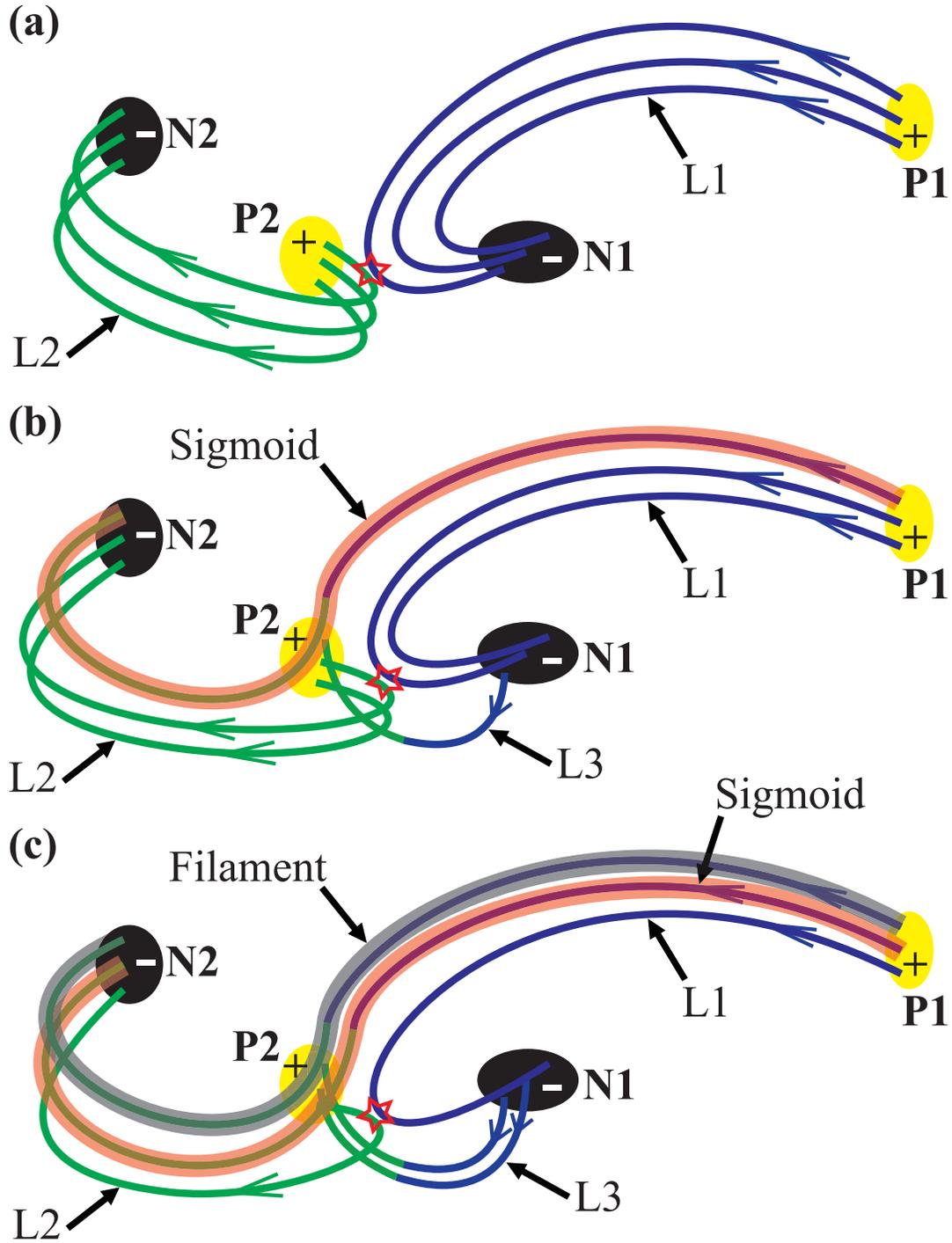}
\caption{Schematic diagrams of the formation of the sigmoid and the subsequent filament.
The yellow (P1 and P2) and black (N1 and N2) ellipses with plus and minus signs represent the positive and negative magnetic fields, respectively.
The green, blue, and green blue lines show magnetic field lines of the loops L1, L2, and L3 and the sigmoid, whose directions are marked by the green and blue arrows.
The red stars denote reconnection between the loops L1 and L2.
The orange and gray thick lines in (b) and (c) indicate plasma separately in the sigmoid and filament.
See Section\,\ref{sec:sum} for details.
\label{f:cartoon}}
\end{figure}

According to the SDO observations, schematic diagrams are provided in Figure\,\ref{f:cartoon} to describe the formation of the sigmoid and the subsequent filament.
Reconnection between the field lines of loops L1 and L2, see red stars at the interface of blue and green lines in Figure \ref{f:cartoon}(a), results in the formation of the newly reconnected sigmoid and loops L3, see Figure \ref{f:cartoon}(b).
Cooling and condensation of plasma in the sigmoid lead to the filament formation, see Figure \ref{f:cartoon}(c).
Along with the successive reconnection, more threads of the sigmoid and the subsequent filament form, see Figure \ref{f:cartoon}(c).

Sigmoid formation by two sets of loops through reconnection is observed.
The flux rope or sigmoid formed via reconnection between two sets of sheared arcades has been reported previously \citep{2008ApJ...683..510K, 2010ApJ...725L..84L, 2015ApJ...812...50J}, that then erupted upward as a CME.
In this Letter, reconnection between two sets of loops is identified, forming the newly reconnected longer sigmoid and shorter loops, see Section\,\ref{sec:res}, in agreement with the tether-cutting model \citep{1980IAUS...91..207M, 2001ApJ...552..833M}.
However, the sigmoid remains stable after its formation, rather than erupts upward as a CME.
This may be caused by weak magnetic twist of the sigmoid.
Similar to \citet{2016ApJ...816...41Y}, the brightening and the increasing EM and temperature are detected in the reconnection region, where the plasma was heated by reconnection. Unlike the case of \citet{2016ApJ...816...41Y}, however, we did not find signatures of the bi-directional expansion of enhanced EM region along the sigmoid. Instead, we observed that the brightenings and regions exhibiting excess EM and temperatures expanded or developed at the eastern endpoint of the sigmoid. This may indicate signatures of material supply into the sigmoid through chromospheric evaporation, caused by reconnection, at the eastern endpoint.
Here, the evaporation flow was not directly detected, e.g., by the Doppler shift, because no spectroscopic data is available. To identify the chromospheric evaporation, the spectroscopic observations are necessary in the future study.
The EM at the reconnection region ($\sim$05:13 UT) increased $\sim$2 minutes earlier than that at the eastern endpoint of the sigmoid ($\sim$05:15 UT). Employing the time interval ($\sim$2 minutes) and the sigmoid length ($\sim$100\,Mm) between these two regions, we suggest that the chromospheric material at the eastern endpoint of the sigmoid was heated by the thermal conduction front and/or beamed non-thermal particles, with a propagating speed of the order of 800 km\,s$^{-1}$, produced at the reconnection region. Alternatively, because the Alfv\'{e}n speed in the corona will be fast enough, an Alfv\'{e}nic perturbation could also have transported the energy from the reconnection region to the endpoints of the sigmoid \citep{2008ApJ...675.1645F}.
The sigmoid with temperature of up to $\sim$2.5 MK is cooler than those identified in the AIA higher-temperature channels, e.g., 94 and 131 \AA, in \citet{2010ApJ...725L..84L}, \citet{2014ApJ...797L..15C, 2018ApJ...869...78C}, and \citet{2015ApJ...812...50J}. 
Although the sigmoid was observed in AIA 94 and 131 \AA~images in this study, it is composed of warm plasma instead.

Filament formation through cooling and condensation of plasma inside a newly formed sigmoid is found.
Coronal condensation, forming a filament, has been presented separately in the loop \citep{liu12} and cavity \citep{berg12}, that show pre-existing structures. 
In this Letter, rather than the pre-existing structure, the sigmoid is newly formed via reconnection between two sets of loops. 
As the sigmoid remains stable after its formation, cooling and condensation of plasma inside the sigmoid are observed, forming the filament.
The transient filament formed through condensation facilitated by interchange reconnection has been reported in \citet{2018ApJ...864L...4L, 2018ApJ...868L..33L, 2019ApJ...884...34L, 2020ApJ...905...26L, 2021ApJ...910...82L, 2021RAA...under...reviewL}, where the condensation occurs in the dips of higher-lying open structures that take part in the reconnection.
In this study, the sigmoid, where condensation happens,  represents the newly reconnected structure via reconnection.
In \citet{2018ApJ...864L...4L, 2018ApJ...868L..33L, 2019ApJ...884...34L, 2020ApJ...905...26L, 2021ApJ...910...82L, 2021RAA...under...reviewL}, the plasma, hardly heated up across the field lines by the lower-lying reconnection, cools down from $\sim$0.9 MK, the characteristic temperature of the AIA 171\,\AA~channel.  
In this Letter, the plasma, supplied from the chromospheric evaporation by reconnection, cools down from a higher temperature of $\sim$2.5 MK.
In situ condensation of coronal plasma, trapped inside the newly formed flux rope that is levitated from the lower corona into the upper corona, is proposed in the reconnection-condensation filament formation models \citep{kane15, kane17, 2021A&A...646A.134J}, in which chromospheric evaporation is not involved. As signatures of chromospheric evaporation are observed in this study, the evaporation-condensation \citep{karp01, 2021ApJ...913L...8H}, rather than the levitation-condensation \citep{kane15, kane17, 2021A&A...646A.134J},  scenario is thus suggested. However, we cannot rule out the levitation-condensation scenario because we do not know if what they proposed is also happening or not in our observations.
All the results support the filament formation models via reconnection and condensation \citep{pneu83}.

\clearpage

\clearpage

\clearpage

\clearpage

\clearpage

\acknowledgments

The authors thank the anonymous referee for helpful comments. We are indebted to the SDO team for providing the data.
AIA images are the courtesy of NASA/SDO and the AIA, EVE, and HMI science teams. 
This work is supported by the Strategic Priority Research Program of Chinese Academy of Sciences (CAS), grant No. XDB 41000000, the National Natural Science Foundations of China (12073042, U2031109,  and 11873059), the Key Research Program of Frontier Sciences (ZDBS-LY-SLH013) and the Key Programs (QYZDJ-SSW-SLH050) of CAS, and Yunnan Academician Workstation of Wang Jingxiu (No. 202005AF150025).
This project has received funding from the European Research Council (ERC) under the European Union's Horizon 2020 research and innovation programme (grant agreement No. 695075).
We acknowledge the usage of JHelioviewer software \citep{2017AA...606A..10M}, and NASA's Astrophysics Data System.


\end{document}